\documentclass[12pt]{article}
\usepackage{epsfig}
\usepackage{amsmath}
\usepackage{amssymb}
\begin{document}

\title{ A note on the time evolution of the fission decay
width under the influence of dissipation}
\author{Helmut Hofmann\\
 \small\it{Physik-Department, T39, TUM, D-85747 Garching,
Germany} }
\date{June 2, 2003}
 \maketitle

\begin{abstract}
The claim put forward in a recent paper by B. Jurado, K.-H.
Schmidt and J. Benlliure that the transient effect  of nuclear
fission may be described simply as a relaxation process in the
upright oscillator around the potential minimum is refuted. Some
critical remarks on the relevance of this effect in general are
added.
\end{abstract}
In the paper \cite{jur-sch-ben} it has been claimed that "a new,
highly realistic analytical approximation to the exact solution of
the Fokker-Planck equation" has been presented. In this note we
should like to raise some questions about the justifications of
the approximations used there, in particular with respect to its
application to the decay of a meta-stable state. Before we address
details of the approach a few remarks of more general nature are
in order.

It has become customary to look at nuclear fission as a time
dependent process. As the current $j_b$ across the barrier shows a
"transient behavior", simply because it takes some finite time
before $j_b$  reaches a quasi stationary value, one likes to
interpret the result in terms of a time dependent decay width
$\Gamma_{\text{f}}(t)$, modifying in this way the one originally
deduced by Kramers using {\em essentially the same picture}. This
transient time seemingly implies a delay of fission during which
light particles might be emitted {\em in addition} to those given
by the conventional ratio $\Gamma_n/\Gamma_{\text{f}}^{\text{st}}$
of the partial widths $\Gamma_n$ for neutron evaporation to the
stationary value $\Gamma_{\text{f}}^{\text{st}}$ for fission, even
if for the latter the Bohr-Wheeler expression $\Gamma_{\text{BW}}$
is replaced by the smaller value $\Gamma_{\text{K}}$ of Kramers'
rate formula. In arguments in favor of such a procedure it is
often claimed that the time dependence comes in only if fission is
considered as a transport process underlying dissipative forces.
In this one forgets that the transition state method is also based
on "collective motion" which, in principle, like particle
emission, is a time dependent event. It is only that for these
processes one has become accustomed to apply widths calculated in
a time {\em in}-dependent picture, in which, in addition, inherent
averaging procedures are applied. Truth is that also Kramers' rate
formula does not represent anything other than an inverse {\em
average decay time}. This can directly be seen by exploiting the
concept of the "mean first passage time" (MFPT). For over-damped
motion an analytic formula for the $\tau_{\text{mfpt}}$ can be
derived from which, under the usual conditions for Kramers' rate
formula, it follows that
$\Gamma_{\text{K}}=\hbar/\tau_{\text{mfpt}}$, see
e.g.\cite{gardiner-STM}. Interestingly enough, the value of the
$\tau_{\text{mfpt}}$ does not depend much on the initial position
of the system, in clear distinction to the transient effect
\cite{HoIv-MFPT-PRL}. In \cite{HoIv-MFPT-PRL} and \cite{HM-MFPT}
the concept of the MFPT has been applied to nuclear fission to
examine, for the limit of over-damped motion, if more light
particles may be emitted than given by the ratio
$\Gamma_n/\Gamma_{\text{K}}$. In \cite{HM-MFPT} it has been
demonstrated that  Kramers' rate formula is valid only for simple
potentials and under favorable conditions for the temperature. For
potentials having some structure in addition to just one
pronounced minimum and one barrier the fission lifetime was seen
to be considerably longer than the
$\tau_{\text{K}}=\hbar/\Gamma_{\text{K}}$ associated to Kramers'
rate. This feature may already by inferred from the
form\begin{equation}\label{tau-smol}\tau_{\text{K}} = \frac{2\pi
\gamma} {\sqrt{C_{\text{a}}|C_{\text{b}}|}}\exp(E_{\text{b}}/T)
\end{equation} the $\tau_{\text{K}}$ takes on for over-damped
motion. Any uncertainty in the product of the two stiffnesses at
the minimum and the barrier, $C_{\text{a}}$ and $C_{\text{b}}$,
respectively, reflects itself in a corresponding error of their
geometric mean, and hence in $\tau_{\text{K}}$. Indeed, these
stiffnesses are known at best in the immediate neighborhood of the
extrema. Realistically, however, the potentials are hardly
symmetric about these extrema. In cases that beyond the top of the
barrier, for instance, the potential becomes wider this property
may effectively imply a smaller $|C_{\text{b}}|$ and, hence, a
larger value of $\tau_{\text{K}}$.

We agree with the authors of \cite{jur-sch-ben} that the
understanding of nuclear dissipation is of great importance, in
particular its variation with shape and temperature. After all,
these are perhaps {\em the} decisive features through which
different models or theories of nuclear transport can be
distinguished \cite{hofrep}. It may perhaps be of interest to
mention that, in addition to the papers cited in
\cite{jur-sch-ben}, quite some work has been done, both
experimentally \cite{pauthoe,dioszegi} as well as theoretically
\cite{hiry}, in which such questions have specifically been
addressed.

The main concern of \cite{jur-sch-ben} is that in previous
analyses of experimental results uncertainties of a factor of two
showed up in the so called reduced friction coefficient $\beta$.
There can be no question that the ultimate goal must be to improve
our understanding about this problem, but it is questionable that
the uncertainties and ambiguities of the method used in
\cite{jur-sch-ben} imply progress. To begin with, one should not
trace all problems back to {\em just the one constant} $\beta$.
Even discarding possible inaccuracies in the height of the
barrier, which enters the decay rate in exponential fashion, there
are crucial problems with the transport coefficients themselves.
The $\beta$, for instance, only stands for the ratio of friction
$\gamma$ to inertia $M$. For obvious physical reasons, these two
quantities must be expected to exhibit a totally different
variation with temperature. Moreover, for a coordinate dependent
inertia, Kramers' formula has to get an additional factor
involving the square root  of the ratio of the inertias at barrier
and minimum \cite{hiry, strutkram}. For truly over-damped motion,
on the other hand, any quantity which involves the inertia looses
any meaning. Indeed, the latter does not appear in formula
(\ref{tau-smol}).

Let us turn now to the more formal problems of \cite{jur-sch-ben}.
The authors aim at delivering a simple way of calculating the time
dependent pre-factor which supposedly relates
$\Gamma_{\text{f}}(t)$ to Kramers' stationary value
$\Gamma_{\text{K}}$. The essential approximation is to calculate
this pre-factor {\em not from a global} solution of Kramers'
equation, which would properly account for the {\em motion across
the barrier}, but from a solution of the same transport equation
{\em restricted to the upright oscillator} by which the fission
potential may be approximated in the {\em neighborhood of the
minimum}. A moments reflection tells one that at the barrier,
where the current $j_b$ is to be calculated, the height of the
artificial potential in this region may easily exceed the barrier
height several times.  The very fact that in this region the
stiffness of this auxiliary potential has the wrong sign is most
crucial for the current, in particular at large times. Whereas for
the inverted oscillator the $j_b(t)$ eventually turns into the
stationary one already found by Kramers, the current for an
upright oscillator tends to zero exponentially. In other words,
replacing already in (\cite{jur-sch-ben}-6) (which is to say in
eq.(6) of ref.\cite{jur-sch-ben}) the correct distribution $W_n$
by the one for an upright oscillator, called $W^{\text{par}}$ in
eq.(\cite{jur-sch-ben}-9), leads to a vanishing denominator in
this basic formula.

To circumvent this problem some intermediate steps are performed
to finally end up with formula (\cite{jur-sch-ben}-8) for which
the $W^{\text{par}}$ of (\cite{jur-sch-ben}-9) is to be inserted.
One basic assumption for this is specified in
eq.(\cite{jur-sch-ben}-7). It implies that, for any time $t$ and
at the barrier top, the dependence of the distribution on
coordinate and velocity is identical to the one at infinite time.
For an oscillator this statement is easily seen to be incorrect,
both for under-damped as well as for over-damped motion (for which
it is claimed to be exact). Take the distribution given in
(\cite{jur-sch-ben}-9), namely\footnote{It is properly normalized,
also in the sense of eq.(\cite{jur-sch-ben}-5) if one only makes
the common assumptions that $x_b$ is sufficiently far away from
the minimum such that the tiny tail beyond $x_b$ does not
influence the normalization integral.}
\begin{equation}\label{dist-par} W^{\text{par}}(x=x_b;t)=
\frac{1}{\sqrt{2\pi}\sigma(t)}
\exp\left(-\frac{x_b^2}{2\sigma^2(t)}\right) \,,\end{equation}
which for an oscillator delivers the correct form for the density
in coordinate space. For over-damped motion this statement is
evident, for under-damped motion one first needs to integrate over
velocity. Putting the form (\ref{dist-par}) into
(\cite{jur-sch-ben}-7)  the $C(t)$ of (\cite{jur-sch-ben}-7) turns
out to be
\begin{equation}\label{ratio-C}
C(t)=\frac{\sigma(t\to\infty)}{\sigma(t)}
\exp\left(-\frac{x_b^2}{2}\left[\frac{1}{\sigma^2(t)}-
\frac{1}{\sigma^2(t\to\infty)}\right]\right) \,,\end{equation}
which evidently is {\em not only a function of time} but varies
with $x_b$. As the upright and inverted oscillator turn into each
other by analytic continuation (changing only the sign of the
stiffness) \cite{hofrep} the proof just given also applies to the
motion of a Gaussian across a barrier, if simulated by a parabola.

For the $\sigma^2(t)$ needed for the $W^{\text{par}}(x=x_b;t)$ of
eq.(\ref{dist-par}) a form is given in eq.(\cite{jur-sch-ben}-10)
which corresponds to zero initial width. On the other hand, the
authors claim it to be more suitable to start from ground state
fluctuations and they try to simulate this feature by
introducing\footnote{It remains unclear why the authors did not
simple generalize the analytic form (\cite{jur-sch-ben}-10) to one
valid for any initial condition, such as (\ref{tdepflu}), shown
below for over-damped motion.} a time shift $t_0$. It is meant to
represent the "time shift needed for the probability distribution
to reach the width of the zero-point motion in deformation space",
which is supposed to be "equal to the time that the average energy
of the collective degree of freedom needs to reach the value
$(1/2)\hbar\omega_1$ associated to the zero-point motion".
Obviously, the authors seem to understand $t_0$ as a kind of
relaxation time to the equilibrium of the oscillator, as
represented by the ground state. It may be noted in passing that
for a genuine quantum system  any application of
(\cite{jur-sch-ben}-10) is prohibited anyway as the distribution
can never have zero width. In any case, it remains unclear why it
should be the ground state and, hence, why there is no influence
of the large intrinsic excitations which are produced in the first
stage of the reaction; after all the authors work with a finite
temperature.

At this stage it may be worth while to remind the reader of some
basic features of transport theory, which may help to clarify a
few critical steps used in \cite{jur-sch-ben}. To begin with, let
us look how quantum features may be accounted for. As it stands,
eq.(\cite{jur-sch-ben}-10) describes relaxation to the equilibrium
specified by the  equipartition theorem of classical mechanics,
represented here by the  pre-factor of the curly bracket. For a
damped oscillator this may be generalized to represent quantum
fluctuations correctly (see e.g. \cite{hofrep}). In equilibrium
they are {\em not} given by those of the ground state
$\hbar/2\mu\omega_1$ used here, for instance in
eq.(\cite{jur-sch-ben}-12). In fact for over-damped motion just
the opposite is true: there the correct quantum equilibrium is
indeed given by the classical limit, see \cite{hofrep} for the
oscillator and \cite{anker-petch} for the general case.

Let us examine now the derivation of eq.(\cite{jur-sch-ben}-12), $
t_0= \hbar\beta/(4\omega_1T)$, meant to determine the time lapse
$t_0$. This equation is obtained by assuming a linear dependence
between the $ \sigma^2(t)$ and time $t$. This approximation is
justified by arguing that the "influence of the potential on the
diffusion process" may be "neglected" as it "is anyhow small in
the range of the zero-point motion". To see the catch in this
argument let us simply write the correct equation for
$\sigma^2(t)$, as it comes out of the Smoluchowski equation for
the oscillator:
\begin{equation}\label{fluc-Smol}\frac{d}{dt} \sigma^2(t) +
2\frac{C}{\gamma} \sigma^2(t) = 2 D_{\text{ovd}} .\end{equation}
Here, $C$ is the stiffness of the potential $U(x)$, such that the
latter may be written as \begin{equation}\label{stiff-cor}U(x) =
\frac{C}{2} x^2 \qquad \text{with}\qquad C=\mu
\omega_1^2,\end{equation} and $D_{\text{ovd}}$ is the diffusion
coefficient, which according to (\ref{fluc-Smol}) is determined by
the equilibrium fluctuation $\sigma^2$ through
\begin{equation}\label{equfudiff} D_{\text{ovd}}= \frac{C}{\gamma}
\sigma^2(t \to \infty)\equiv\frac{C}{\gamma}\sigma_{\text{eq}}^2
\qquad \approx \frac{T}{\gamma} .\end{equation}
Eq.(\ref{fluc-Smol}) implies that the "influence of the potential
on the diffusion process" is given by the second term on the left
{\em which has the same size independent} of the "range" of the
coordinate. The solution of eq.(\ref{fluc-Smol}) for $t\ge 0$ is
given by
\begin{equation} \label{tdepflu} \sigma^2(t)=
\Bigl(\sigma^2(t=0)-\sigma_{\text{eq}}^2\Bigr)\exp\left(-\frac{2C}{\gamma}
t\right)+ \sigma_{\text{eq}}^2 ,\end{equation} showing that
relaxation to the equilibrium value happens on the time scale
$\tau_{\text{ovd}}=\gamma/2C$ independent of the initial
fluctuation. Of course, for $t \ll \tau_{\text{ovd}}$ and zero
initial fluctuations the $\sigma^2(t)$ becomes linear in $t$, but
the reason why such a $\sigma^2(t)$ should be identified as the
{\em ground state fluctuation} of the {\em undamped} oscillator
remains unclear. Put in the context mentioned above: it is unclear
why such a value of $\sigma^2(t) \equiv \hbar/2\mu\omega_1$ should
be relevant if {\em reached by a process of strong damping}.

Next we turn to the value found for $t_0$ from
eq.(\cite{jur-sch-ben}-12). It turns out so small that the
introduction of this quantity and the associated fluctuation
cannot explain why the $\Gamma_f(t)$ starts to become finite only
at about $0.7\cdot 10^{-21}$ s. Indeed, for $\beta=2\cdot 10^{21}
\text{s}^{-1}$, $\hbar \omega_1 = 1$ MeV and $T=3$ MeV one gets
the very small number of $t_0\simeq 0.06\cdot 10^{-21}$ s. The
explanation given in  \cite{HoIv-MFPT-PRL} comes much closer:
There, by simulating the whole fission process by a Langevin
equation,  it was demonstrated that such a shift is related to the
relaxation of the initial distribution to the quasi-equilibrium in
the minimum. For the numbers just used the relaxation time
$\tau_{\text{ovd}}$ becomes $\tau_{\text{ovd}}\simeq 0.36 \cdot
10^{-21}$ s --- provided one makes use of the relation
(\ref{stiff-cor}). The value of $\tau_{\text{ovd}}$ becomes even
{\em very} close to the $\simeq 0.7 \cdot 10^{-21}$ s at which in
Fig.\cite{jur-sch-ben}-1 the $\Gamma_{\text{f}}(t)$ is seen to
rise if the relation of frequency to stiffness is replaced by the
{\em in}-correct one given in equation (\cite{jur-sch-ben}-13)
where the stiffness $K$ is assumed to be only half the correct
value given in (\ref{stiff-cor}), in accord with the common
definition used in text books not only on nuclear physics but on
classical and quantum mechanics as well. It is true that, for the
cases discussed in Fig.\cite{jur-sch-ben}-1, the motion is not
really over-damped (for $\hbar \omega_1 = 1$), but the
$\tau_{\text{ovd}}$ may nevertheless be taken as a fair estimate.

As indicated before, it is left unclear why in the general case
the system should start with a small fluctuation. Indeed, formula
(\cite{jur-sch-ben}-8) together with (\cite{jur-sch-ben}-9)
implies any transient effect (of the type discussed here) to be
absent if one chooses to start out of the quasi-equilibrium, which
is to say for $W^{par}(x=x_b,t=0)=W^{par}(x=x_b,t\to \infty)$. For
a bound system, like the upright oscillator, such an initial
condition implies that the system stays in equilibrium for ever.
For over-damped motion this may be seen from eq.(\ref{tdepflu})
together with (\ref{dist-par}). For a meta-stable situation like
fission, on the other hand, the situation is different: then there
will be a finite current outwards. Exactly this feature is not
described correctly by formulas (\cite{jur-sch-ben}-8) and
(\cite{jur-sch-ben}-9).  Please recall that in many cases an
initial condition like that of the quasi-equilibrium specified
before is not at all unrealistic. For sufficiently large fission
barriers, as they are required for Kramers's rate formula anyhow,
the system may well have enough time to reach such a stage around
the first well before it decays by fission.

Let us finally comment on the feature that for certain cases the
present construction seems to represent fairly well the
numerically obtained global solutions of the transport equation
for the full fission potential. In our opinion this feature should
be considered accidental rather than supply a decent basis for
trustworthy applications in future work of the approximations
advertised in this paper. There are simply too many
inconsistencies to warrant applicability to the general case.


\begin{thebibliography}{99}
\bibitem{jur-sch-ben} B. Jurado, K.-H. Schmidt and J. Benlliure,
Phys. Lett. B 553 (2003) 186
 \bibitem{gardiner-STM} C.W. Gardiner, "Handbook of stochastic methods,
   Springer, 2002, Berlin
\bibitem{HoIv-MFPT-PRL} H. Hofmann and F.A. Ivanyuk,
 Phys. Rev. Lett.90.132701
\bibitem{HM-MFPT} H. Hofmann and A.G. Magner,  to appear in PRC,
see also nucl-th/0304022
\bibitem{hofrep} H. Hofmann, Phys. Rep. 284 (4\&5) (1997) 137-380
\bibitem{pauthoe} D.J. Hofman, B.B. Back, I. Di\'oszegi,
C.P. Montoya, S. Schadmand, R.Varma, and P. Paul, Phys. Rev.Let.
{\bf 72}, (1994) 470; see also:  P. Paul and M. Thoennessen,
Ann.Rev.Part.Nucl.Sci. {\bf 44}, (1994) 65.
\bibitem{dioszegi} I. Di\'{o}szegi, N.P. Shaw, I. Mazumdar, A.
Hatzikoutelis and P. Paul, Phys.Rev. C {\bf 61}, (2000) 024613.
\bibitem{hiry} H. Hofmann, F.A. Ivanyuk, C. Rummel and S. Yamaji, Phys. Rev.C,
         64 (2001) 054316
\bibitem{strutkram} V.M. Strutinsky, Phys. Lett. B {\bf 47}, 121 (1973).
%%%%%%%%%%%%%%%%%
\bibitem{anker-petch} P. Petchukas, J. Ankerhold and H. Grabert,
 Annalen der Physik (Leipzig)(2000)  1;
J. Ankerhold, P. Petchukas and H. Grabert, Phys. Rev. Let. 87
(2001) 086802

\end{thebibliography}
\end{document}